\documentclass[%
 reprint,
 amsmath,amssymb,
 aps,
]{revtex4-1}

\usepackage{graphicx}
\usepackage{dcolumn}
\usepackage{bm}

\usepackage[T1]{fontenc}	
\usepackage[utf8x]{inputenc}

\usepackage{geometry}		
\usepackage{graphicx}
\usepackage[above,below]{placeins}	
\usepackage{times}
\usepackage{float}
\usepackage{natbib}
\begin{document}

\title{Cooperation is not enough: The role of instructional strategies in cooperative learning in physics education.}

\title{Undergrad classroom cooperation and academic performance: Beneficial for real-world-like problems but detrimental for algebra-based problems}



\author{J. Pulgar}
 \altaffiliation{Departamento de Física, Universidad del Bio Bío, Chile. \\ jpulgar@ubiobio.cl}
\affiliation{
 Departamento de Física, Universidad del Bío Bío, Concepción, Chile.
}%


\author{Cristian Candia}
\homepage{Kellogg School of Management, Northwestern University, Evanston, IL 60208; Centro de Investigación en Complejidad Social (CICS), Facultad de Gobierno, Universidad del Desarrollo, Las Condes, Chile. \\ cristian.candia@kellogg.northwestern.edu}
\affiliation{
Kellogg School of Management, Northwestern University, Evanston, IL 60208
}%
\affiliation{
Northwestern Institute on Complex Systems (NICO), Northwestern University, Evanston, IL 60208
}%
\affiliation{
Centro de Investigación en Complejidad Social (CICS), Facultad de Gobierno, Universidad del Desarrollo, Las Condes, Chile.
}
\author{P. Leonardi}
\affiliation{%
Technology and Management, College of Engineering, University of California Santa Barbara, CA.\\
}%

\date{\today}

\begin{abstract}

For several decades, scholars have studied cooperation and its outcomes in the educational context. Yet, we lack a complete understanding of how different instructional strategies impact the relationship between cooperation and learning. Here we studied how different instructional strategies led to different social configurations and their differences in individual academic performance in an experiment with 82 first-year students from an introductory physics course. Surprisingly, we found that students who actively seek out information on multiple peers are less likely to achieve good performance on well-structured (algebra-based) problems, whereas, for ill-structured (real-world-like) problems, this effect depended on the features of the learning environment. Besides, we observed that good performance on ill and well-structured problems responded to different social network configurations. In a highly clustered network (which contains redundant information), students performed well-structured problems better than ill-structured problems. In contrast, students with access to network structural holes (which enable access to more diverse information) performed ill-structured problems better than well-structured problems. Finally, ill-structured problems could promote creative thinking, provided that instructors guide the solving process and motivate students to engage in the appropriate cognitive demands these problems entail. Our results suggest that teaching and instructional strategies play an important role in cooperative learning; therefore, educators implementing cooperative learning methods have to accompany them with adequate instructional strategy.

\end{abstract}

\maketitle


\section{\label{sec:level1}Introduction}

Cooperation is fruitful for learning \citep{Johnson_Johnson, johnson1989cooperation} leading to better academic outcomes in the educational context \citep{Kassarnig2018, Blansky2013, Sacerdote2011, Biancani, Candia}. Moreover, cooperation among peers eases collective learning, and the emergence of good ideas, processes linked with crucial competencies in today's society \citep{pllegrino}.  Yet, little is known on how different instructional strategies impact the relationship between cooperation and learning.

Here, we investigated students' social networks from three different classes of an introductory physics course and determined the social structures that facilitate good performance on well-structured physics problems (e.g., algebra-based problems), and an ill-structured physics problem \citep{Jonassen}. By well-structured problems, we refer to algebra-based problems, often characterized in Physics Education Research (PER), by simplified and idealized situations that have little to no connection with students' real-world experience \citep{Heller_Hollabaugh}, and frequently found on physics textbooks \citep{Chi, Larkin}. On the other hand, by ill-structured problems, we refer to those problems associated with real-world problems \citep{Fortus} that lack the information that individuals would use to find an already known and unique solution. Ill-structured problems introduce high levels of uncertainty associated with a spectrum of possible outcomes and strategies on how to proceed to create them \citep{Shin, Jonassen}. Concretely, here we use a task consisted on student groups generating physics problems for high school students. 

Through this study, we explored whether different forms of collaborative mechanisms --creative combinations (CC) \cite{Burt2004, Burt2005, Burt2016, Hardagon} or interrogation logic (IL) \cite{Rhee}-- predict good performance on well and ill-structured problems. To this aim, we set three different experimental conditions for 82 first-year students, where we varied teaching and instructional strategies. The experiment was run in an introductory physics course over two months at a university in Northern Chile. To explore the extent to which students' social structures facilitated academic performance, we collected data on students' performance on a physics test designed upon well-structured problems, and performance on an ill-structured problem. In addition, we asked students to respond to an on-line peer-nomination survey related to their social interactions engaged for information seeking to solve problems. Finally, we tested the effects of different instructional strategies on academic performance and whether collaboration responded differently depending on the learning environment.

\subsection{\label{sec:level2}Physics Problem Solving and Collaboration}

Well and ill-structured problems have different characteristics that might enact different forms of collaboration. In physics education, algebra-based problems (i.e., well-structured tasks) demand the use of a limited number of rules and principles (e.g. algebra and physics principles), along with a set of procedures that are well organized, constrained to certain parameters (e.g., initial and/or the final conditions on a motion problem in kinematics). Besides, these tasks have predictable actions that are frequently used to solve similar problems \citep{Dufresne,Kohl2008}. Good performance on algebra-based problems has not been reported necessarily as a consequence of conceptual understanding \cite{Kim,Byun}, as students tend to solve such tasks through a 'plug and chug' strategy \cite{Dufresne, Larkin, Byun}. In addition, well-structured problems can be defined as disjunctive tasks \cite{Steiner1966} with low levels of positive interdependence \cite{Johnson_Johnson}, as these tasks might be solved by the most capable or vocal students when addressed in groups. 

On the other hand, the difficulty of ill-structured problems relies on deciding the appropriate constraining conditions that would guide solvers to transition the open-ended scenario towards decisions to come up with  their unique response \citep{Rietman}. Fortus \cite{Fortus} studied the importance of making assumptions when solving ill-structured mechanics problems on experts and novices. He found that even experts struggled for creating adequate assumptions on the physics variables and principles involved, and on the absolute or relative magnitudes of the variables for deciding and developing solutions. From the embedded attributes of ill-structured problems, one might expect ill-structured problems to introduce high levels of positive interdependence \cite{Johnson_Johnson}, and be perceived as additive tasks \citep{Steiner1966}, where performance emerged as the sum of all members’ contributions and relevant abilities \cite{Dreu2011}. These expectations are coherent with the experience from Heller and colleagues \cite{Heller_Keith}, who designed context-rich problems as an alternative to traditional textbook physics activities \cite{Heller_Hollabaugh}, and found that groups performed better than isolated students.

\subsection{Network Centrality and Learning}
Social Network Theory provides two alternative collaborative mechanisms that enable knowledge development and idea generation for problem-solving: a. creative combinations (CC);  and, b. interrogation logic (IL),  where both collaborative mechanisms are oriented towards the emergence of good ideas but through different social configurations. The former, CC of information is a mechanism that depends on one's structural position within the network, and the diversity of knowledge that could be accessed through such structure. Accordingly, good ideas would depend on how people learn new information through different social ties, from zones of high knowledge redundancy (i.e., high network cohesion), to zones where actors have access to isolated partitions of the network (i.e., structural holes). Actors who bridge connections between two unconnected individuals or groups, or who span structural wholes through brokerage, would enjoy the advantages of social capital by accessing the resources available in different places of the network and are therefore more likely to produce creative ideas \citep{Burt2004, Fleming}. In addition, individuals and groups located in central positions of the social network are more likely to have creative outputs, because they are placed in paths connecting two or more teams, and therefore have access to the information that is transferred through those links \citep{Tsai, pentalnd, Burt2004, Fleming}. In contrast, peripheral students and groups placed at the end of the information path would depend on central groups letting the knowledge flow in their direction, thus making them less likely to take faster advantage of the information flowing throughout the network \citep{Dawson}. 

Differently, the IL \cite{Rhee} is a mechanism where highly constrained networks (e.g., cohesive groups) afford opportunities for creative ideas, but though different cognitive processes than actors who span structural holes. On tasks that benefit from IL, subjects' attention is focused on specific content and its related ideas rather than on the diversity of information flowing throughout the network, which enabled them an in-depth examination of the local knowledge managed by the individuals embedded in the cohesive group.

In Education and PER literature, we found no reference in regards to whether the aforementioned collaborative mechanisms (i.e., CC and IL) are preferred for solving different types of problems. So far, education researchers have used network analysis to explore the academic advantages of a central position in students' networks. Academic performance is most likely to be enhanced by being immersed in a cohesive social network from which students can take advantage of the information, skills, abilities others might share through social ties \citep{Gasevic, Smith, Borgatti_Cross, Reagans, Fleming, Calvo-Armengol2009}. Research evidence has found significant correlations between centrality measures and performance \cite{Putnik, Bruun2013}. Moreover, the teaching and learning conditions play an essential role in encouraging (hindering) student social interaction for students to reach central positions in the network \cite{pulgar, Brewe_Kramer2012}. Finally, recent evidence has found that the number of social ties (e.g., centrality) is not a straightforward predictor for academic achievement. More out-ties for cooperation showed a negative effects on students' performance, while reciprocal ties led to better academic performance \cite{Candia}, adding a new condition over the nature of the social relationship for academic achievements. 

From this body of evidence, we asked: What is the role of instructional strategies (teaching and learning activities) in cooperation and learning outcomes in an undergraduate physics course?

\section{\label{sec:level41}Methods}

We conducted an experiment in three undergraduate sections from an introductory physics course designed for engineer majors in a University in Northern Chile, over a period of 8 weeks in 2018. We aimed to explore whether student collaboration played either similar or different effects over performance on well and ill-structured physics problems. For this purpose, in collaboration with course instructors, we designed a battery of ill-structured problems grounded on real-life situations for a weekly administration during problem solving sessions for a period of 7 weeks. 

Students were engineering majors in their first or second year of college education, pursuing a careers on either Industrial Civil Engineer or Software Civil Engineer. A total of 82 participated in the study (Traditional = 33; Mixed = 23; Treatment = 26). 

\begin{figure*}
  \includegraphics[scale=0.5]{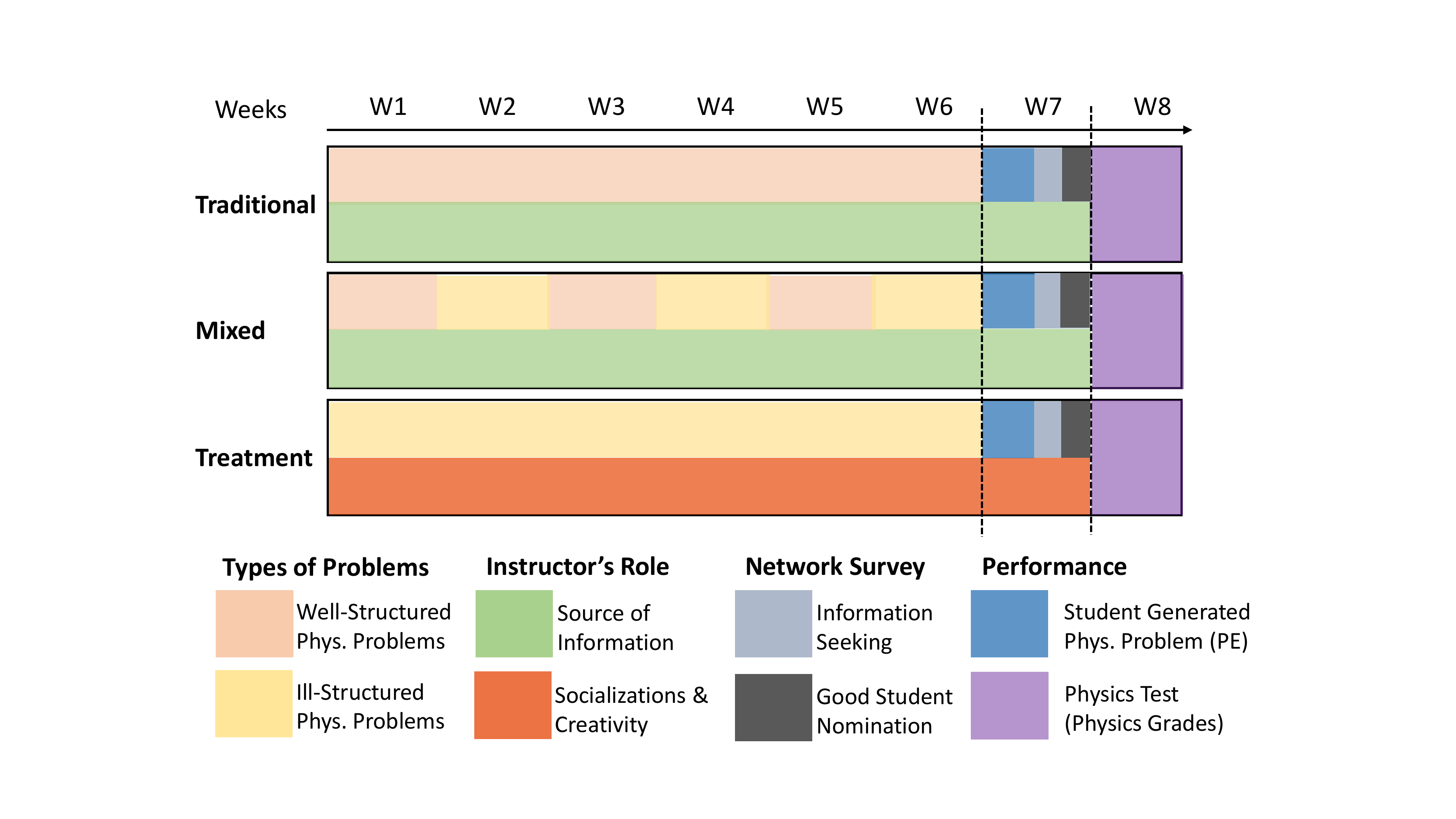}
  \caption{Experiment design, with three sections: Traditional, Mixed and Treatment. The diagram shows the timeline of the events from weeks 1 through 8. Includes the unique instructional characteristics per section (Types of Problems, and Instructor's Role); along with Performance instruments and Network Surveys. \label{Experiment}}
\end{figure*}

The details of the experiment are depicted in Fig. \ref{Experiment}. Here, we show the instructional characteristics of the three sections (Traditional, Mixed and Treatment), the types of physics problems they worked, and the role each instructor enacted in guiding the sessions. In terms of instructor's role, we identified two alternative behaviors: a) Source of information, that is, instructor facilitated direct information to respond to students' questions regarding the problem; and b) Socialization and Creativity, that is, instructor responded to students questions by directing their attention to other classmates who may have either asked similar question, or responded it already. Fig. \ref{Experiment} also shows performance and network instruments administered on weeks 7 and 8. More details of these instruments in the following section. 

\subsection{\label{sec:level22}Data Collection}
During the 7th week of the experiment (Fig. \ref{Experiment}), we tasked students with the activity of designing a physics problem for high school students (see the activity instructions in supplementary material) addressing the concepts and principles of circular motion. At the end of the session we gathered students' generated physics problems, and asked them to respond an online peer-nomination survey to identify the social network of the class during the problem solving session where participants solved the ill-structured problem.

\begin{table*}[t!]
\begin{center}
  \caption{Code description of problem characteristics for Problem Elaboration.\label{Tabla2}}
 
\begin{tabular}{p{5cm} p{11cm}}
\hline\hline
Code                                    & Description \\
\hline 
Physics Concepts Asked                  & Physics concepts used as problem items (e.g., angular speed, tangential acceleration).\\
\textit{Type of Information}            &                                               \\
\hspace{0.2cm}Ready-to-Use Info         & Data is explicitly presented in the problem and with appropriate units for its use.\\
\hspace{0.2cm}Conversion of Units       & Physical quantities that need conversion to respect the IS of units (i.e., m and s).                                    \\
\hspace{0.2cm}Text to Math              & Physics information is presented in written form and needs translation into mathematical expressions (e.g., ``begin its motion from rest'' or ``uniform motion''). \\
\hspace{0.2cm}Algebra Transformation    & Physics information for solving the problem needs algebraic steps for accessing and using it.  \\
\hspace{0.2cm}Information Research      & The problem requires researching appropriate magnitudes to solve the problem. \\
\hspace{0.2cm}Assumptions	            & Problem forces students to assume particular characteristics of the problem, such as constant acceleration, or the position of the ‘particle’ that describes the circular motion.\\
No. Phys. Concepts Asked                & Number of physics concepts used as problem items. \\
No. Equations Needed                    & Number of equations required to solve the problem.\\
Contextual Details                      & Elements from real-life activities, and/or actors witnessing or engaging in actions. \\
Word Count                              & Number of words used on the problems' description.\\
Cognitive Demand                        & Taken from a taxonomy of introductory physics problems \citep{teo}.\\
\hline
\end{tabular}
\end{center}
\end{table*}

\subsubsection{Network Surveys}

The survey consisted on two questions administered through Qualtrics online survey service that aimed to measured the following networks (see survey design on supplementary material): \paragraph{} 

Network of Information Seeking: From whom had you sought information for solving the physics problem addressed in this session?
\paragraph{}

Network of Good Students: Who is a good physics problem solver in your class? (i.e., a student you believe is good at understanding physics content and solving physics problems).

To facilitate students’ responses on each of these questions, we included the roster of students enrolled per section. Consequently, subjects responded by selecting the individuals in their sections from whom they sought information, and the ones that are perceived as a good students. Both questions led to directed (i.e., ties are not necessarily reciprocal) and binary networks (i.e., links between nodes either exist (1) or do not exist (0)). The network of information seeking was designed to reveal whether students engaged on social interactions with the goal of finding resources and ideas for solving the physics ill-structured problem. Because flow ties are difficult to obtain, social interactions such as 'seeking information' may be perceived as proxies of information flow \cite{Borgatti, pentalnd}. Using good student network is thought to enable an additional dimension to reveal what type of students engaged on information seeking, to then explore whether this perceived prestige is a valuable contributor to the social processes that affect academic success.

\subsubsection{Dependent Variables}

Instructors of the course facilitated the variable Physics Grades, which consisted on students' scores to a test designed by instructors over three algebra-based problems administered on week 8 (Fig. \ref{Experiment}), and used in the analysis as the measurement of performance on well-structured physics problems. Physics grades were shared by the instructors three weeks after the day of data collection, without the possibility to review the assessment instrument, nor students’ solutions to these problems. 

Finally, the performance measurement for students solutions to ill-structured problems is labeled as Problem Elaboration (PE), and was constructed to assess the degree of elaboration in students' generated problems for the content of kinematic of circular motion. The following is an example of a problem generated by one student group: 

\begin{quote}
    Donkey Kong wants to throw barrels to King K Rool. For this, Donkey Kong throws one barrel with an angular speed of 2$\pi$ rad/s. By knowing that at 3 s its speed is 10$\pi$ rad/s, and that the barrel impacts at 5 s. Determine: a. the angle covered by the barrel; b. the magnitude of the centripetal and tangential acceleration at the moment of impact at 6 cm from its center; and c. frequency and period. 
\end{quote}

Because creative tasks and their respective outputs may deviate from the standard solutions, analyzing performance on ill-structured problems was conducted through the identification of embedded features and characteristics. We conducted the analysis on a total of 26 problems (Traditional $=$ 10; Mixed $=$ 9; Treatment $=$ 7). In order to conduct the analysis, we translated these problems from Spanish to English, which were revised by a native English speaker knowledgeable in physics. The analysis of these solutions (i.e., physics problems) was conducted on NVivo 12 plus, a software for qualitative data analysis. This qualitative description came from the identification problems’ attributes and characteristics, such as physics concepts used as data and/or questions, type of information, contextual details, word count among others variables shown in Table \ref{Tabla2}. A first wave of problem coding was conducted by the lead author, which yielded to an initial version of the code-book, who was revisited in collaboration with a trained graduate student in qualitative analysis and physics content (For more details see \cite{PulgarPERC}). After agreement, an independent wave of coding was performed, where both covered $40\%$ of the data (10 problems), obtaining a Cohen's Kappa of .92.

\subsubsection{Network Measures}

\begin{figure}
  \includegraphics[width=1\linewidth]{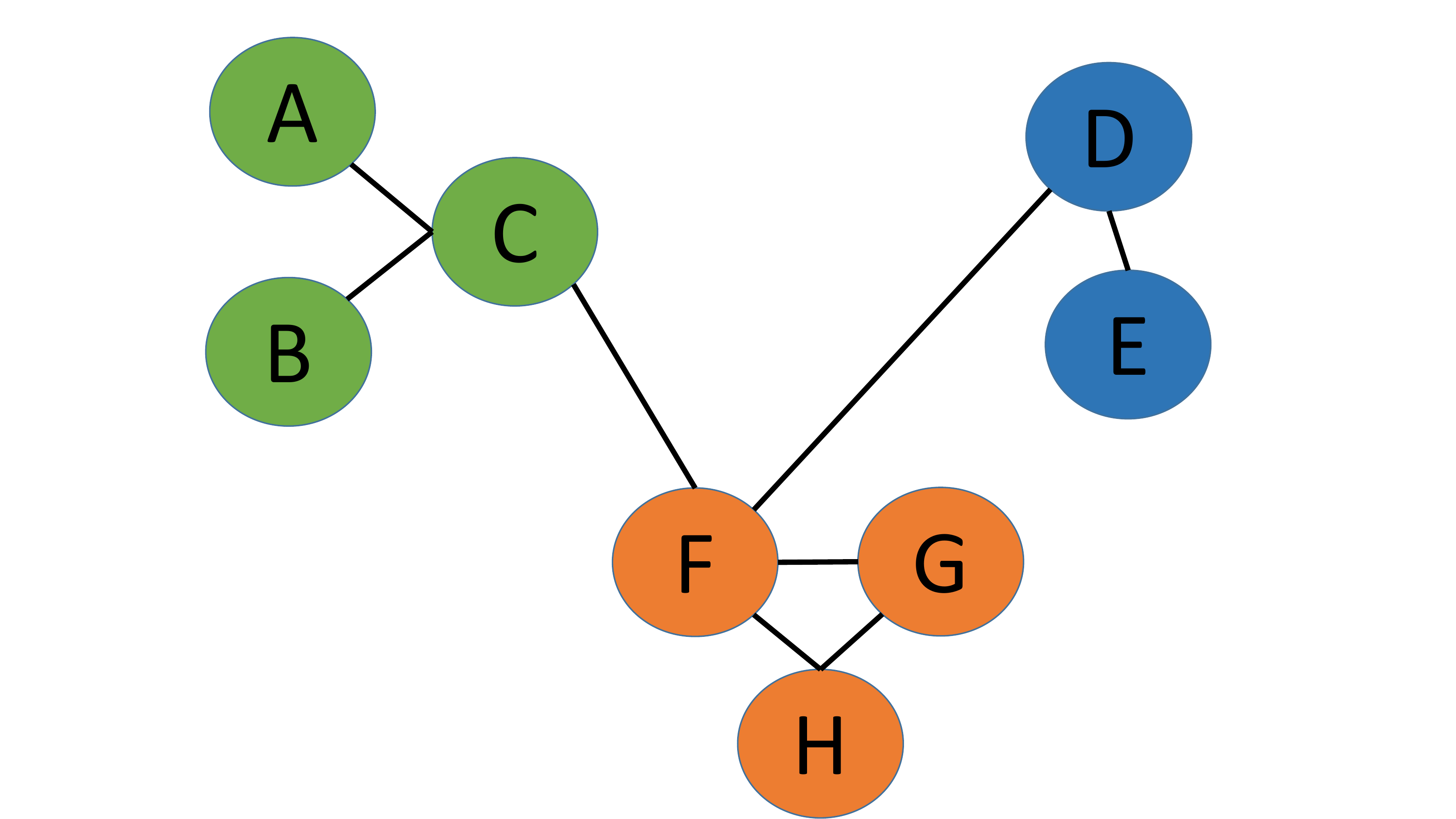}
  \caption{Network diagram of constraints and structural holes. Node $F$ has access to different sources of information from blue and green communities. \label{Figure1}}
\end{figure}

The network measures used for this analysis were computed from the network of information seeking (i.e., response to survey question a). This set of social structure variables consisted on different metrics of network centrality (degree, in-degree, out-degree, betweenness  and eigenvector ), as well as network constraint and a brokerage metric known as gatekeeper. Following we describe each of these variables:  

\begin{itemize}
  \item \textit{Degree} is a network measure of centrality that counts the number of edges (i.e., social ties) connecting the focal actor. 
  
   \item \textit{Out-degree}: on directed networks this measure of centrality counts the number of outgoing edges or social ties for a given node, that is, the number of links directed from the focal actor towards other individuals within the network.
   
  \item\textit{Gatekeeper}: a brokerage measure that counts the number of times node $i$ bridged connections between $j$ and $q$, being the source node $j$ a member of a different group than $i$ and $q$, which in turn are members of the same group. A gatekeeper broker is an individual that spans non-redundant ties with nodes outside its own group, has connections with its own group members, and engages in bringing information from the outside ties, while the destination of that information is a members within its own group. On Fig. \ref{Figure1}, nodes C, D and F display such type of brokerage as they display ties with nodes outside their own units (sources), but at the same time engaged with teammates, and therefore, may have access to novel information from these outside sources and bring it to the group.
  
  \item\textit{Eigenvector}: network centrality measure that regards to social influence within a system, as it depends on whether the nodes tied to the focal actor shows social ties to other well connected nodes.
  
  Accounting for the connectivity of one's friends is key for flow processes \citep{Borgatti}, to the extent that friends with social relationships outside one’s social domain might boost chances of receiving and sharing valuable information for learning, innovation and social status. The algebraic representation of eigenvector is as follows:$e_i$=$\lambda\sum_{j}x_{ij}e_{j}$. Here, $e_i$ $i$s the eigenvector centrality of node $i$, and $\lambda$ the largest eigenvalue of $e_i$. Moreover, $x_{ij}$ can take values of 1 or 0 depending on whether nodes $i$ connected to $j$ or not respectively. That is, eigenvector centrality of node $i$ is proportional to the sum of its neighbors’ eigenvector centralities.
  
  \item\textit{Constraint} Constraint is network measure that accounts the number of redundant social ties, that is, the degree to which a node spans ties with others who are also connected to each other \citep{burt1992}.
  
  This is an inverse measure of brokerage, or the node that bridges isolated portions of the network, thus accessing structural holes. High constraint will indicate that a node is totally invested in a group of already connected others, and will therefore have access to zero structural holes. The definition introduced by \cite{burt1992}: $C_{i}$=$\sum_{j}c_{ij}$ ,$i$$\neq$$j$; $c_{ij}$=$(p_{ij}+\sum_{q}p_{iq}p_{jq})^2$,$q$$\neq$$i$,$j$, where $C_{i}$  is the constrain of node $i$, and $c_{ij}$ an index that indicates $i$’s investment on its relationship with $j$, counting direct ($p_{ij}$: proportion of tie strength between $i$ and $j$, relative al all of $i$’s ties) and indirect ($\sum_{q}p_{iq}p_{jq})$: proportion of tie strength through indirect paths connecting $i$ and $j$ via $q$). 
  
  Network constraint is a variable negatively associated with brokerage, that is, the investment in social interactions that bridge connections between previously isolated portion of the network, known as structural holes. For instance, on Fig. \ref{Figure1}, node F has access to a structural hole because it shows non-redundant ties between groups green and blue, and may access to new information and ideas from both groups, which may provide unique opportunities for creative combinations. Consequently, node F would have lower network constraint than, for instance, nodes G and H as these have redundant ties, and therefore are unable of brokering beyond their close network. 
\end{itemize}

\subsubsection{Control Variables}

Finally, we accessed to data on students' scores on a nationwide standardized testing (University Selection Test or UST) to access higher education, type of high school from where students graduated, city where they lived before entering university, engineering major and gender, which were utilized as control variables in our analysis. These control variables aim to account for the homophily mechanisms that drive social networks configuration in higher education. \cite{candia2019higher, Biancani, Kassarnig2018}.

\subsection{\label{sec:level23}Data Analysis}
After removing missing cases, the number of students remaining for analysis was N = 67. We used ordinary least square multiple regressions (OLS) on the continuous dependent variables (i.e., physics grades and problem elaboration) to explore the effect of network structures, as well differences in performance by sections. 

First we tested the effect of network measures over problem elaboration and physics grades. For this we regressed physics grades on network predictors. The models for grades and problem elaboration include interaction terms between class sections and the investigated network measure, which enable a comparison and interpretation on whether the network variable has a similar effect over the whole sample, or its effect over students’ outcomes depends on the learning environment defined by the type of problems and teaching strategy. In order to ease interpretation of regression coefficients, all predictors were standardized. For interpreting the regression coefficients of categorical variables such as academic sections (as), school type (st) and engineer major (em), readers must consider that the coefficient emerges as the difference between the variable in the model and the baseline categories (here as: Traditional, st: Industrial civil engineer, and  em: public schools). 

Later, we explored whether engaging on problem elaboration enabled good performance through the moderation of social engagement on information seeking. In other words, we investigated the degree to which creative problems foster students' ability to answer well-structured problems in interaction with students' network structure. For this purpose, we fitted OLS multiple regression models with an interaction term between problem elaboration and network measures.

\section{\label{sec:level24}Results}

\subsection{\label{sec:level26}The Effect of Social Structures over Physics Grades}

\begin{figure*}
 \centering
 \includegraphics[scale=0.3]{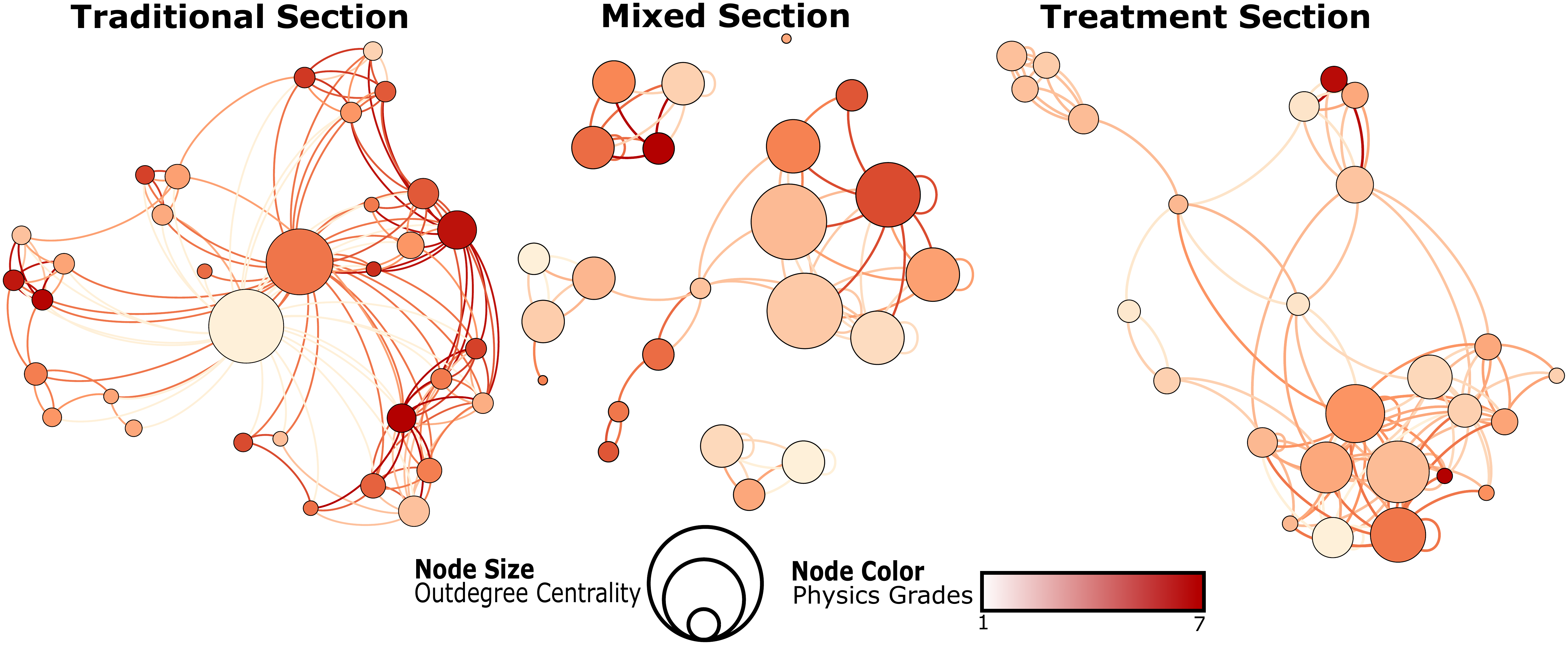}
 \caption{Classroom networks for the three analyzed sections: Traditional, Mixed, and Treatment. Node color represents physics grades (being dark red the highest), and the node size represents the out-degree centrality, i.e., the number of times that a student seeks for information in the classroom.\label{Figure3}}
\end{figure*}

\begin{figure*}
 \centering
 \includegraphics[scale=0.3]{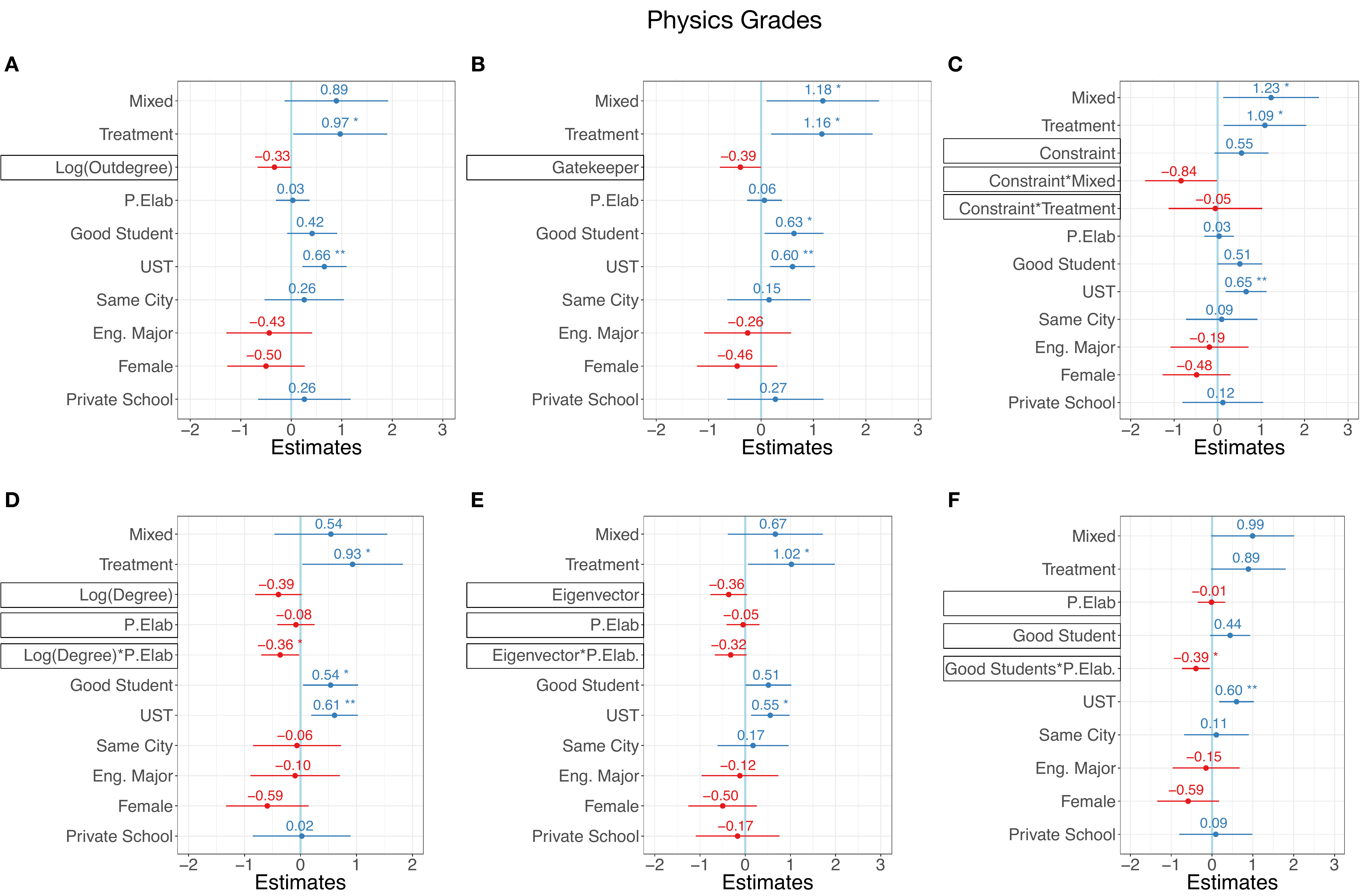}
 \caption{Graphic depiction of OLS multiple regression models for Physics Grades regressed on network predictors, controlling by confounding variables (see table 2 on the supplementary results section). Red color indicates a negative effect and blue color indicates a positive effect. One star (*) indicates significance at a level of p-value $<0.05$ and two stars (**) indicate significance at a level of p-value $<0.01$.\label{Model_Grades}}
\end{figure*}

Figure \ref{Figure3} depicts classroom networks for information seeking. The node size represents out-degree centrality --the number of times that the student seeks for information to a peer,-- whereas color shades indicate the grade obtained in the physics test (well-structured problem). By visually inspecting the figure, we observe high grades students --nodes with darker colors-- tend to be smaller (i.e., lower out-degree) and located at the periphery of the network. In contrast, low grades students tend to have higher out-degree.

Figure \ref{Model_Grades} summarizes the multiple regression models fitted using log(out-degree) (\ref{Model_Grades}A), gatekeeper (\ref{Model_Grades}B),  network constraint (\ref{Model_Grades}C), log(degree) (\ref{Model_Grades}D), eigenvector (\ref{Model_Grades}E), and the baseline of good student centrality (\ref{Model_Grades}F). These models allowed us to explore the effect of network structures over physics grades, and whether such effects are invariant of the teaching conditions enacted on each section. The regression coefficient for the Treatment section is positive and significant, with a large effect over physics grades in all models, even after controlling for all the confounding variables. Students under the Treatment condition are likely to increase almost a point in their grades, compared to what students in the Traditional section would score under similar conditions. This result suggests important effects of the learning environment generated in the Treatment section, based on ill-structured problems, along with guidance over socialization of information. 


Surprisingly, and contrary to research evidence in the literature, centrality metrics showed a negative effect over grades. These effects are observed for log(out-degree) (A), gatekeeper (B) , log(degree) (D) and eigenvector (E). Because out-degree refers to the number of outgoing ties, the activity of seeking out information does not afford good grades. In general, having a high number of social ties, either incoming or outgoing, shows to be negatively related to physics grades, as seen on model D for log(degree). Consistent with our previous results, connecting others outside one's group for information seeking (Figure \ref{Model_Grades}B) does not afford academic success in well-structured problems.    

\begin{figure}
 \centering
 \includegraphics[width=0.85\linewidth]{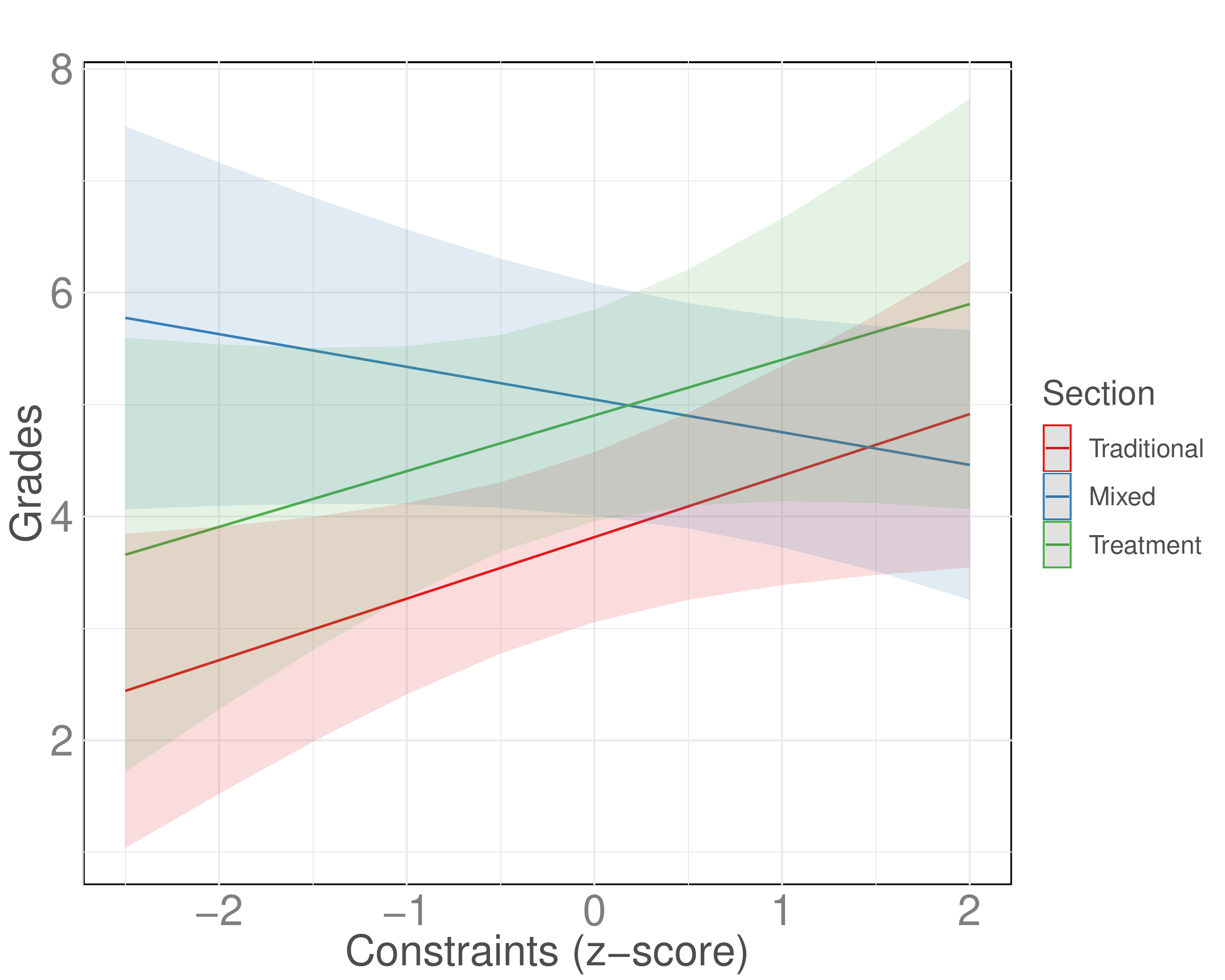}
 \caption{Interaction between network constraint and sections for predicting physics grades. \label{Figure4}}
\end{figure}

Different from other models, the regression coefficient for network constraint in \ref{Model_Grades}C is positive, yet not statistically significant. The direction of the coefficient is consistent with the collaborative process of interrogation logic \cite{Rhee}, a process that benefits from highly constraint networks. To disentangle this relationship, Figure \ref{Figure4} shows the interaction between network constraint and classroom sections in predicting physics grades. According to the plot, both Traditional (red) and Treatment sections (green) show positive and statistically the same slopes, whereas the effect of network constraint is negative for the Mixed section (blue). This result suggests that high access to structural holes (i.e., low constraint) leads to higher grades just for the Mixed section.


\subsection{\label{sec:level25}The Effect of Social Structures over Problem Elaboration}

\begin{figure*}
 \centering
 \includegraphics[scale=0.3]{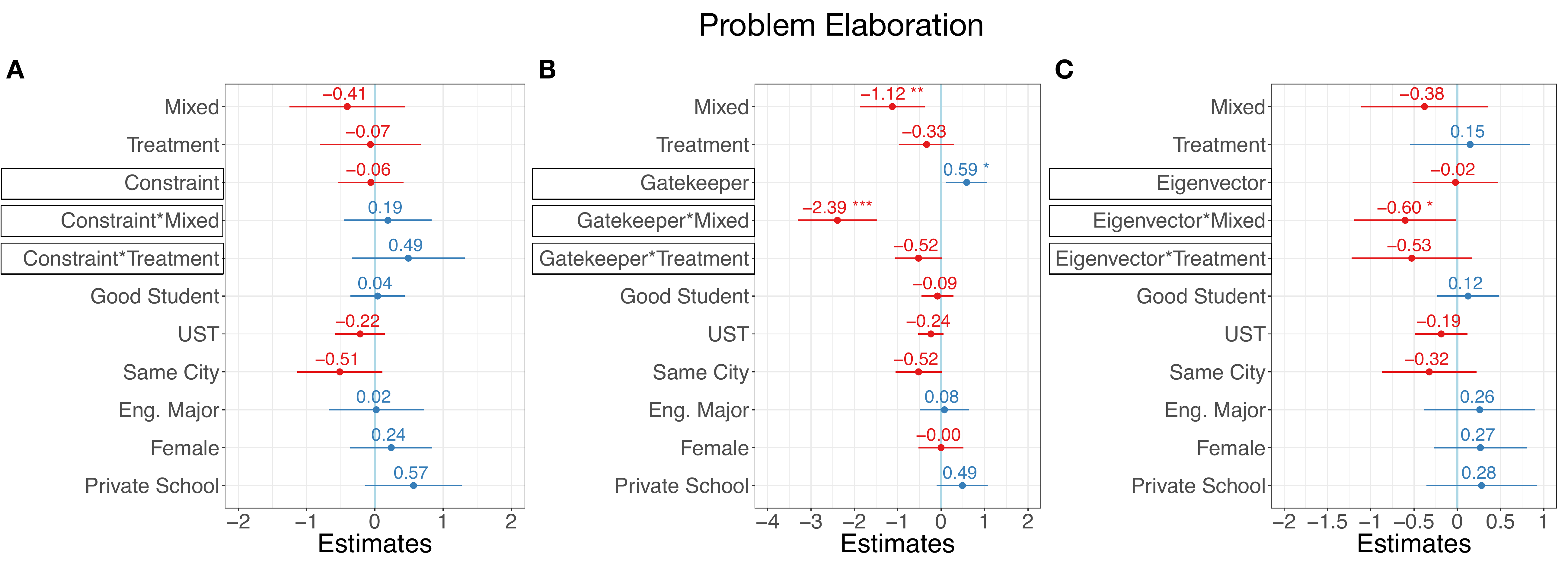}
 \caption{Graphic depiction of OLS multiple regression models for Problem Elaboration regressed on network predictors, controlling by confounding variables (see table 1 on the supplementary results section). Red color indicates a negative effect and blue color indicates a positive effect. One star (*) indicates significance at a level of p-value $<0.05$ and two stars (**) indicate significance at a level of p-value $<0.01$. \label{Models_PE}}
\end{figure*}

Figure \ref{Models_PE} summarized the multiple regression models on problem elaboration (See Supplementary Table 1), with main predictors in academic section, network constraint (Fig. \ref{Models_PE}A), gatekeeper brokerage (Fig. \ref{Models_PE}B), and eigenvector centrality (Fig. \ref{Models_PE}C). Also, we include the interaction between the network metric and the academic section variable to explore whether there are differences in problem elaboration due to differences in instruction. For all models, we control for different confounding variables such as good student nomination; higher education application score (UST); a dummy variable that takes value $1$ if students reside in the same city as their family; a dummy variable gender; and a dummy variable for private or non-private high school.

Figure \ref{Models_PE}A shows that network constraint --being a member of a cohesive network with redundant ties-- has a positive but non-significant effect over problem elaboration.

\begin{figure}[!t]
 \centering
 \includegraphics[width=0.9\linewidth]{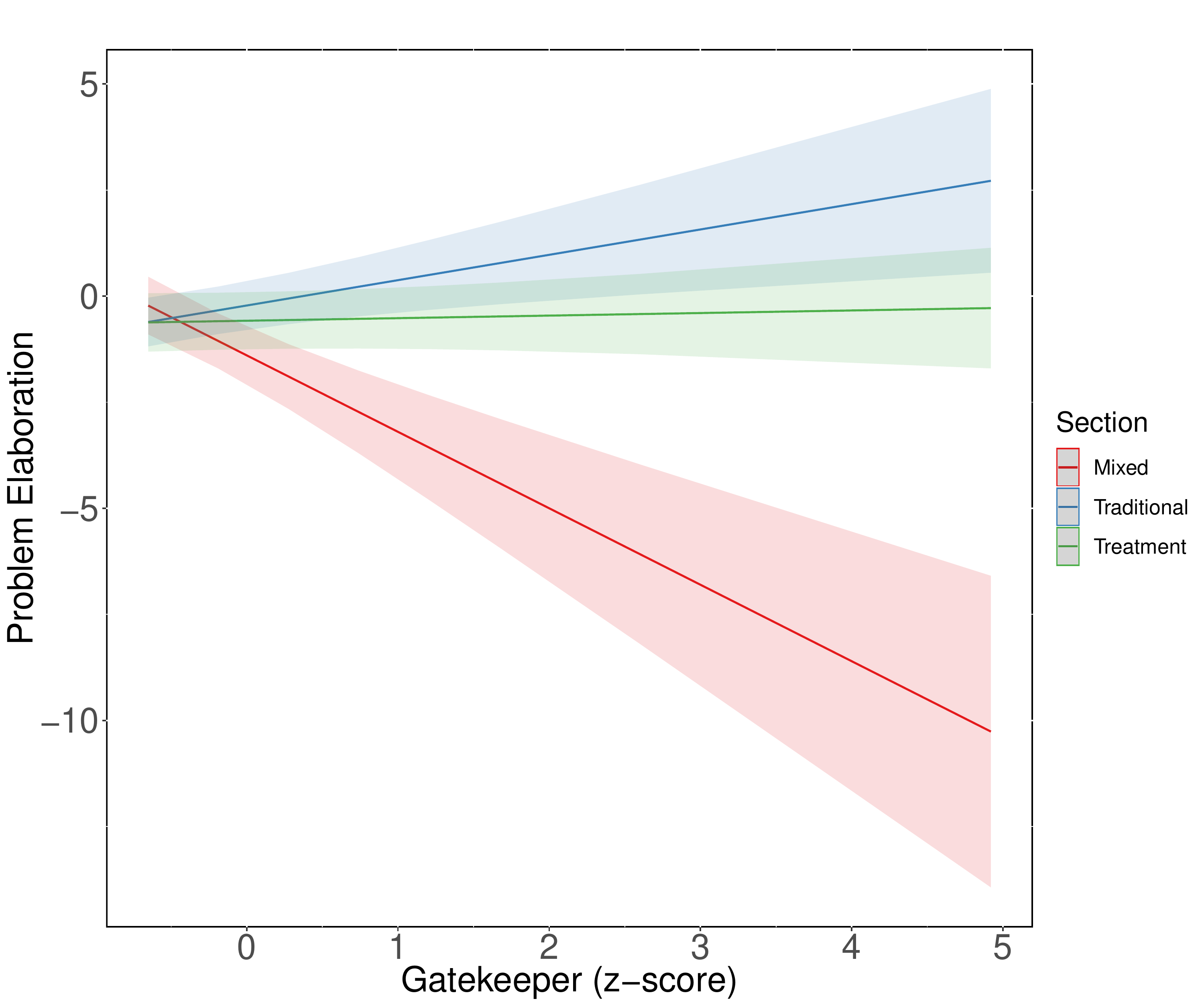}
 \caption{Linear regression for interaction between gatekeeper and sections for predicting physics grades.  \label{Figure2}}
\end{figure}

Figure \ref{Models_PE}B shows a negative and significant difference in problem elaboration between the Mixed and Traditional section, while such difference is also negative but not significant between Treatment and Traditional. Moreover, students seeking out information from peers from other groups, and sharing it with their team members (i.e., gatekeeper brokerage) is a positive predictor of problem elaboration above and beyond instructional differences. The interaction term is negative for Mixed compared Traditional section and statistically significant at $.001$ level and less negative for Treatment relative to Traditional, but at $.1$ level of significance. Figure \ref{Figure2} depicts the relationship between problem elaboration and gatekeeper by section. Accordingly, the Mixed section exhibits a negative slope, while being a gatekeeper in Traditional and Treatment sections yield to higher problem elaboration.

Figure \ref{Models_PE}C shows that eigenvector centrality to be negatively related to problem elaboration. In other words, students who are linked to well-connected others in the network of information seeking perform worse in problem elaboration than students who are not well-connected. However, the effect size gets closer to zero when including the interaction between eigenvector and classroom sections, which means that just Mixed and Treatment sections mainly drive the previously described effect. 

\subsection{\label{sec:level27}The Moderation Effect of Network Structures}

Here we explore whether network variables drive the relationship between problem elaboration and physics scores. Because having designed an elaborated physics problem had no direct effect in predicting physics grades (Fig. \ref{Model_Grades}), we considered the possibility that this relationship to be moderated by students’ structural position within the network of information seeking. Here we present multiple regression models with moderators in log(degree) and eigenvector centrality (Fig. \ref{Model_Grades}D and E, respectively). Besides, we tested the moderating effect of perceived good students over the relationship between problem elaboration and physics grades (Fig. \ref{Model_Grades}F).  We follow the rationale that different levels of problem elaboration may have enabled differences in conceptual understanding and abilities for solving well-structured problems (i.e., physics grades), at different levels of perceived status (i.e., good students). Both models, Figures \ref{Model_Grades} D and E, showed negative interaction terms between network centralities and problem elaboration in predicting grades. We found the same result for the moderated effect of good student nomination, with a negative coefficient.

\begin{figure*}
 \centering
  \includegraphics[width=0.4\linewidth]{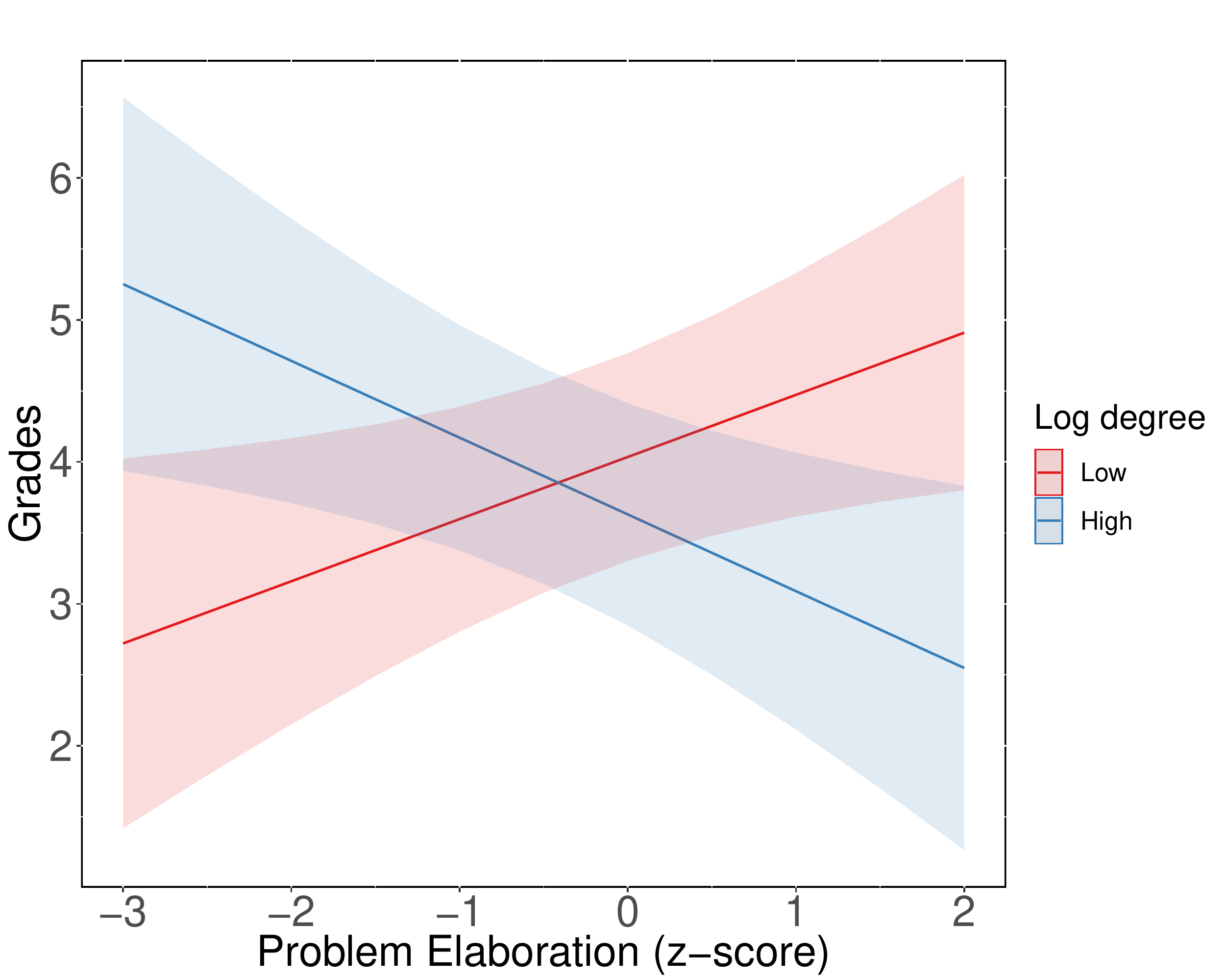}
  \includegraphics[width=0.4\linewidth]{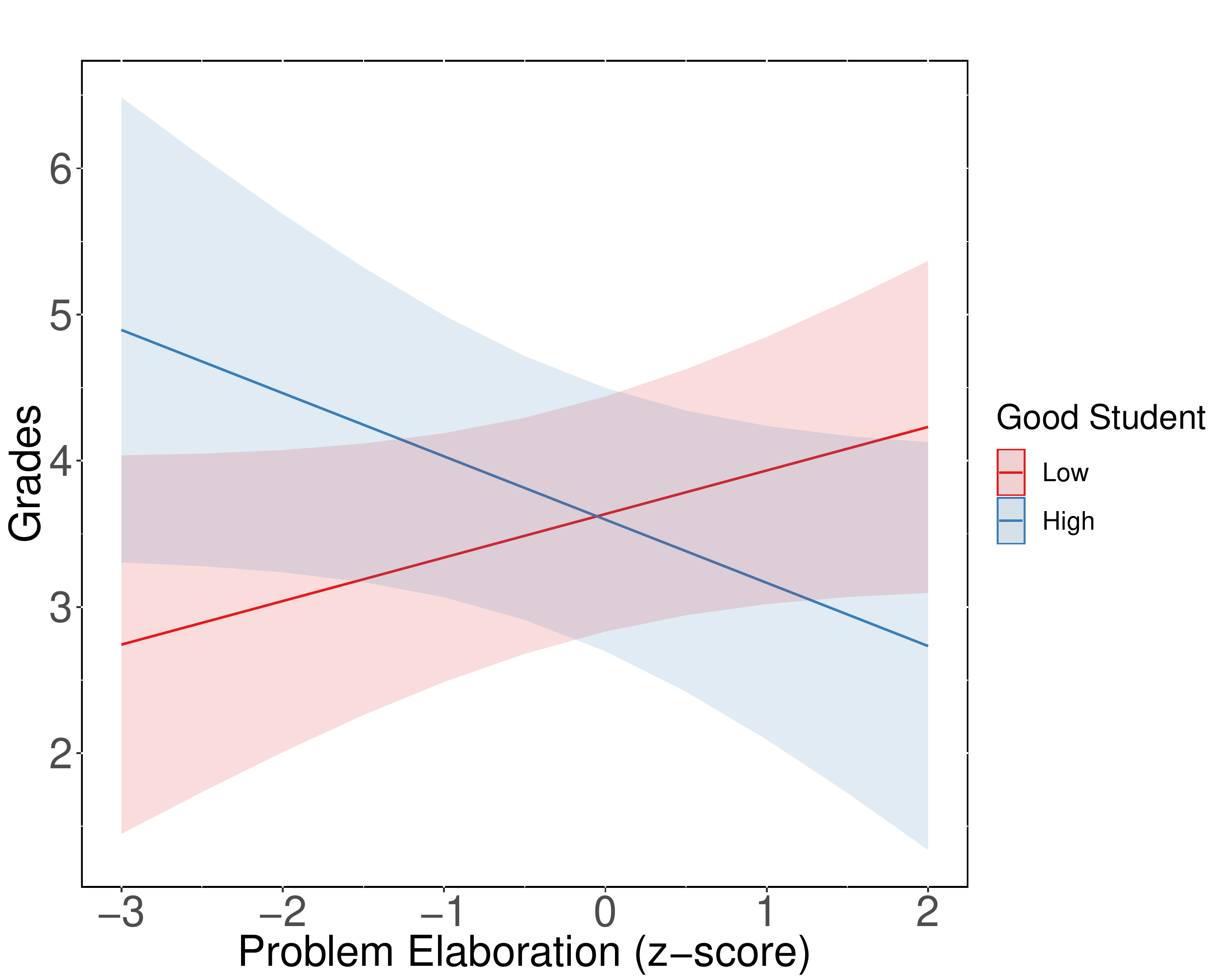}
  \label{fig5}
  \caption{Interaction between problem elaboration and A) the logarithm of node degree, and B) good student nominations, both from a linear model predicting physics grades. The shaded areas represent a level of confidence interval of $0.9$ and low (high) intervals contains the values below (above) the mean of each variable. \label{Figure5}}
 \end{figure*}

Fig. \ref{Figure5} depicts the relationship between problem elaboration and physics grades at different levels of log(degree) and good students nomination. First, students who showed low degree centrality (red, Fig. \ref{Figure5}A) benefited from developing problems with high elaboration, as this process afforded them good grades. However, for students with high degree centrality (blue), creating highly elaborated problems showed a detrimental effect over grades. In simple words, scoring high in problem elaboration enabled good grades only for those who engaged in less social interactions for information seeking (i.e., low log(degree)), which is coherent with our previous results. Finally, Fig. \ref{Figure5} B depicts the relationship between problem elaboration and physics grades at different levels of good students nomination. The interaction plot shows that participants who were not perceived as good students in physics (red) benefited from creating well-elaborated problems, which in turn translated in good grades, while 'good students' (blue) might have gotten better grades after creating problems with low levels of elaboration. 

\section{\label{sec:level42}Discussion}


Based on the types of problems worked on the Mixed section, it is a surprise that the Mixed section had lower elaboration than the Traditional section (see \ref{Models_PE} model B). The learning conditions, problems and instructional guidance on how to solve problems on each section may have influenced students’ motivation for creating problems with various levels of elaboration and complexity. For instance, the learning goal of the task (i.e., design a physics problem for secondary students) may have motivated students in the Traditional section to utilize characteristics from textbook problems that were in their repository of activities to design problems in an effective way. The Mixed section worked on ill-structured problems, but the instructor did not emphasize the importance of assumptions in the face of ill-structured activities. Consequently, highlighting the role of assumption making when addressing creative tasks we believe had positive effects over students' expectations and motivation for generating problems, as suggested by the high problem elaboration found on problems from Treatment section, whose instructor engaged in such a positive narrative for creativity. 

Interestingly, being a central actor within the network of information seeking does not afford good grades. This evidence is was observed for variables such as outdegree, degree and eigenvector centrality, and consistent with the evidence found in Candia and colleagues \cite{Candia}. The directionality of this relationship contradicts the research evidence found on other studies in Education and PER \cite{Putnik,Bruun2013,Grunspan}. To understand this contradictory results, one could focus on the nature of the social networks mapped on this and other studies, and argue in favor of the nature of the social processes as one of the reasons why we obtained contradictory evidence. Studies in PER had asked students to write down the names of their peers with whom respondents had meaningful interactions inside the classroom \citep{Williams,Zwolak, Bruun2013}. Under such survey question, students are likely to remember interactions with friends \citep{eagle}, or useful interactions related to the learning goals of the session \citep{Bruun2013}. Differently, the survey question used in this study aimed to determine students’ social engagement in the process of seeking out information in the classroom, where students were also likely to report useful as well as friendship-based interactions for information seeking. However, both types of relationships may not necessarily overlap as the nature of the network does not account for the effectiveness of the social tie (i.e., whether the information accessed was useful or not). That is, students may have interacted and reported ties with friends and others not consider friends for information for solving the problem, regardless of the meaningfulness of the interactions. Consequently and according to the negative coefficients of centrality over physics grades, students are either not capable of requesting appropriate information for solving physics problems due to ineffective communication, or it may be that engaging in such processes for information seeking is irrelevant in the learning context described here. If the former were true, this would be evidence for the need to engage students on the social processes linked to effective communication and collaboration. Yet, if the learning context were blind to social interactions and sharing information, then this would call for a reflection over the teaching and learning practices involved in university education. Alternatively, it may be the case that students approximated effective social interactions, yet the actors reached lacked meaningful information to share, or rather provided misconceptions regarding the content and/or the goals of the task. Consequently, having nodes with reduced knowledge of the content is not an ideal scenario for students to engage in socialization of information for collective growth. This calls for remedial strategies that prepare subjects for proper learning before putting them in positions to collaborate.

It is worth paying attention to the significant interaction between network constraint and sections for predicting physics grades (see \ref{Figure4}). Here, both Traditional and Treatment show a positive relationship with grades, whereas for Mixed section this relationship is negative. This evidence suggest that the social systems created under Traditional and Treatment conditions take advantage of highly constrained networks, where subjects presumably engaged on deep analysis and reflection of ideas, or interrogation logic \cite{Rhee}. Consequently, within such a cohesive network it is easier to learn complex information, as well as to develop good ideas \citep{Fleming}. This process is evidence that the nature of well-structured problems does not benefit from the mechanism of creative combinations, but rather engaging on such efforts brings negative effects. Access to unique connections is related to inflow of novel ideas, which here does not afford better outcomes, likely because the well-bounded nature of the physics information for solving well-structured problems does not need novelty, but rather conventional knowledge. Further, the negative effect of constraint on the Mixed section suggests the opposite, where students benefit from connecting structural holes. Surprisingly, students on the Mixed section displayed higher network constraint relative to students from Traditional section (Fig. \ref{Figure2}). Consequently, not taking advantage of it for scoring higher grades may be due to ineffective communication for collaboration. 

Moreover, and even though the models did not yield to significant coefficients, constraint showed null effect for problem elaboration compared to the negative physics grades, while gatekeeper brokerage showed to be a positive predictor for problem elaboration, and negative for physics grades. These results add interesting evidence to the contrasting nature of both types of performance, as well as the shape of learning objectives and the measurement instruments design for such purpose. Generating problems may be close to benefiting from creative combinations \citep{Burt2004} compared to well-structured physics problems, provided students engaged on effective mechanisms for information seeking in a context that rewards creativity like the Treatment section. 

The moderated effect of network centrality and good student nomination for predicting physics grades are consistent with the single effect of network structures over physics grades. These results constitute additional evidence of the detrimental effect of socialization and seeking out information, presumably through ineffective mechanisms. Surprisingly, students who are not perceived as good students would get better grades if they score higher on problem elaboration. Alternatively, the complexity of generating a physics problems showed to have negative effects for students who enjoy the social recognition of being proficient in physics. The physics education tradition grounded on algebra-based physics problems \citep{Byun,Chi,Kim,Larkin}, and its consequent belief that a good physics performance responds to solving well-structured problems has clearly encouraged students to recognize proficient others based on their ability to solve such tasks. However, the set of skills to solve algebra-based problems may not necessarily enabled them better outcomes in more creative-oriented tasks. This evidence challenges the nature and features of proficiency in this particular context, and pushes us to expand our own perspectives in the matter. 



\section{Limitations and Future Recommendations}
We recognize the limitations of this study associated with the reduced sample size, and the lack of alternative variables that would have strengthen the analysis of students’ responses and social experience. Further control and observation over instructional strategies would also facilitate a deeper understanding of the nature of the social system generated on each academic section. In addition, short term activities for a single session might discourage interdependency and continues collaboration among students. Consequently, future pedagogical innovations should include higher level of structure, with explicit learning goals at individual and group level. An important dimension for improvement consists on understanding the different ways in which students collaborate and gain access to information from their peers. Such effort might support the interpretation that students engaged on ineffective forms of communication when solving different types of activities, which would lead to recommendations over the importance of appropriate strategies for social capital depending on the nature of the task. Additional evidence on students’ strategies for social connection may support the need to introduce pedagogical innovations that respond to creativity and collaboration in university education. Based on this results, educators must be cautious in implementing teaching strategies grounded on principles of collaboration and interdependency. Using such principles would demands intense attention on students’ interactions, and appropriate guidance over effective strategies for collaboration and communication of information. In addition, introducing ill-structured problems in education brings positive learning outcomes and interesting chances for creative thinking, yet, this is true when instructors guided the solving process and motivated students to engage on the appropriate cognitive demands these problems entail. In addition, having students developing appropriate content knowledge before attempting to introduce activities that require intense knowledge transfer may induce richer dialogues.

\section{\label{sec:level43}Conclusions}

Encouraging students to solve open-ended activities such as ill-structured problems within a learning environment that highlights the importance of creativity and socialization of information (i.e., Treatment condition) would likely boost students' opportunities for better grades compared to traditional classrooms. Yet, the process of interacting with others for the purpose of accessing new information for solving ill-structured problems might be detrimental for obtaining good physics grades, particularly if the latter performance measurement comes as the results of algebra-based problems (i.e., well-structured tasks). The nature of the well-structured problems and the features of the learning context tend reward individualized performance, or collective efforts that emerged from highly cohesive clusters of students.  Moreover, well and ill-structured problems responded positively to different social structures, and therefore, social positions that afforded good grades may have unwanted effects if the tasks are ill-structured. Finally, to optimize the learning effects of socialization of information, the learning environment must enable appropriate distribution of knowledge, encourage and value social interactions and the emergence of unconventional ideas, as well as effective communication.

\section*{Data Availability}
The anonymized data necessary to reproduce this work can be delivered under a reasonable request to the authors.

\section*{Code availability}
The entire analysis and data processing were done using the standard R libraries (https://www.r-project.org/).

\section*{Author Contributions}
J.P. contributed to the study design, acquisition and data analysis and writing the manuscript. C.C. contributed with the data analysis, study conception and writing the manuscript. P.L. contributed to the study conception and writing the manuscript.

\section*{Corresponding Authors}
Javier Pulgar (jpulgar@ubiobio.cl).\\
Cristian Candia (cristian.candia@kellogg.northwestern.edu | ccandiav@mit.edu | ccandiav@udd.cl). \\

\section*{Competing interests}
Authors declare no competing interests.

\acknowledgments{This material is based upon work supported by the AAPT E. Leonard Jossem International Education Fund. Any opinions, findings, and conclusions or recommendations expressed in this material are those of the author(s) and do not necessarily reflect the views of the American Association of Physics Teachers.}

\bibliographystyle{unsrt}
\bibliography{ref}

\pagebreak
\clearpage
\section*{Supplementary Material}

\subsection{Experimental Protocol}

Students enrolled in their respective sections through on-line University system. Yet, they had no knowledge regarding the name of the instructor responsible for the section they selected. The introductory physics course addressed the contents of kinematics and newtonian physics. The course had three 2 hours sessions each week, plus a laboratory practice session. Two of these sessions were dedicated to lectures, while the remaining one was defined as a problem-solving practice session. The intervention with ill-structured problems was administered during the problem-solving sessions, where instructors on the Mixed and Treatment sections distributed problems for students to work in groups. Students self-selected their teams.

\begin{figure}[!h]
\centering
  \includegraphics[width=1\linewidth]{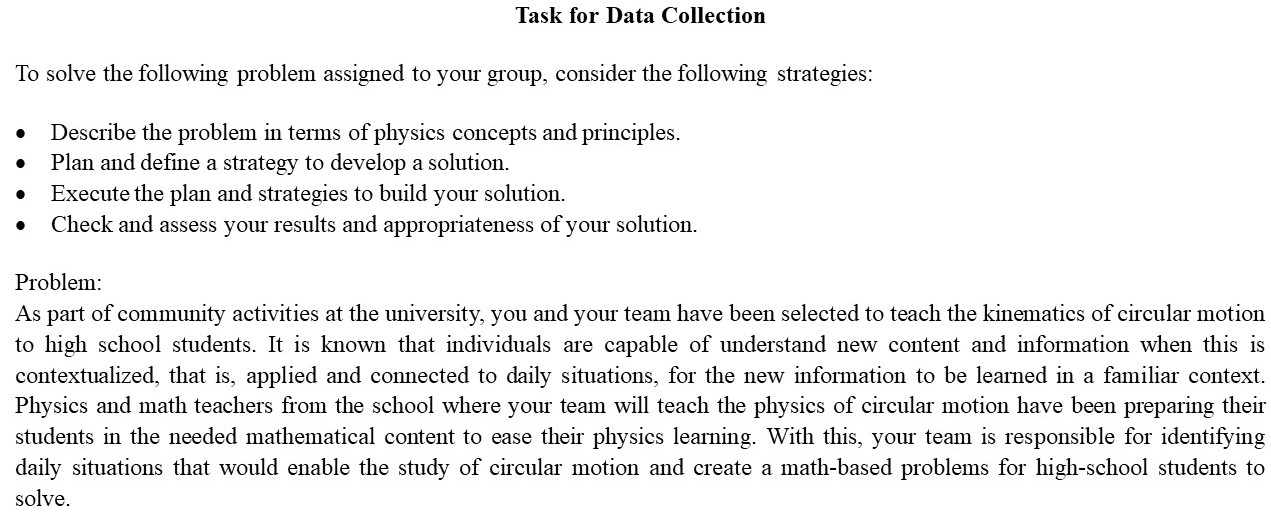}
  \caption{Ill-structured problems administered to course participants. \label{Ill-structured}}
\end{figure}

\begin{figure}[!h]
    \centering
    \includegraphics[trim=245 60 100 60, clip,width=1.3\linewidth]{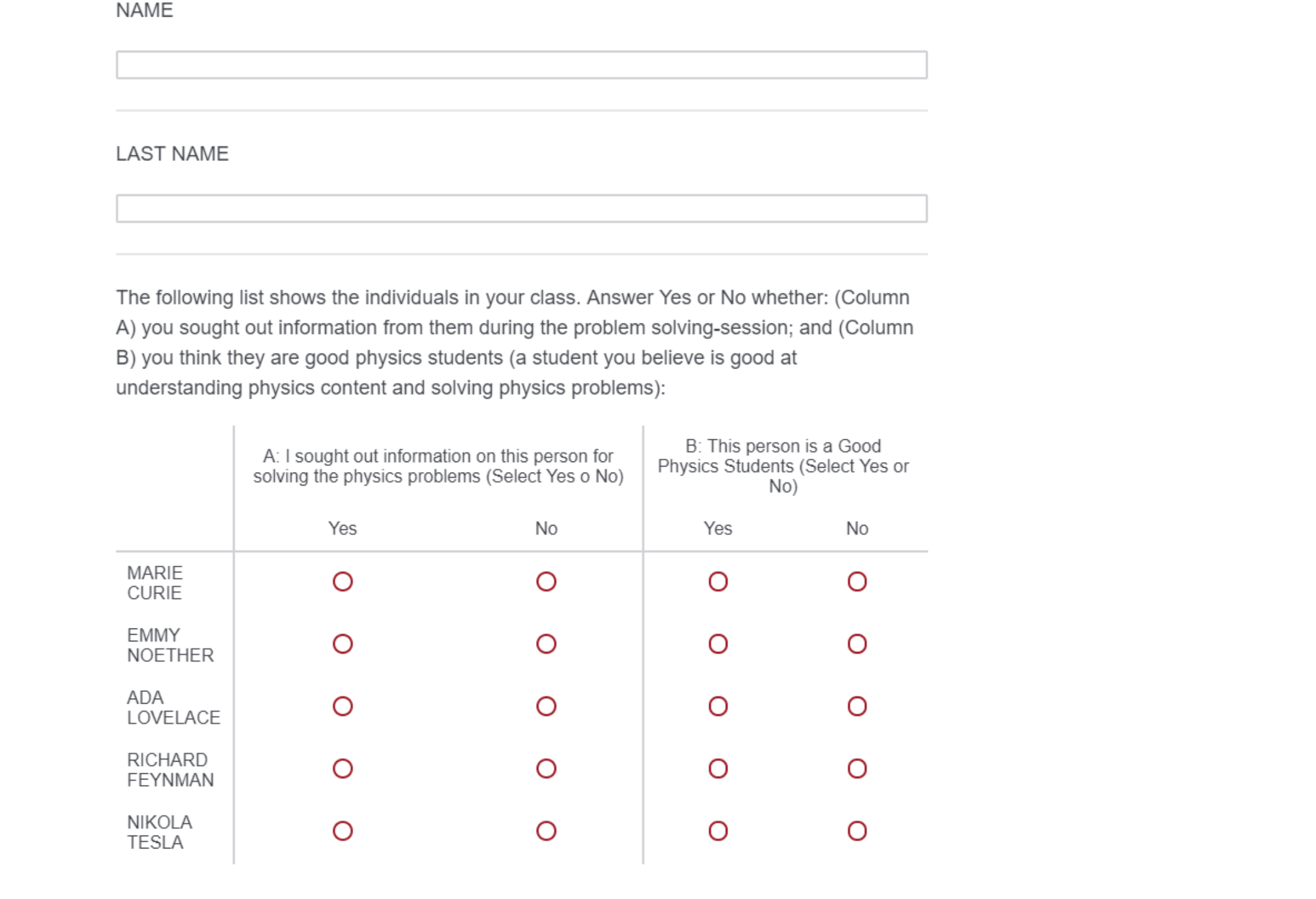}
    \caption{Survey instrument consisting on two questions: (A) Network of Information Seeking; and (B) Network of Good Students.}
    \label{survey}
\end{figure}

On week 7, we tasked students with the activity shown in Fig. \ref{Ill-structured}, where we asked to design a physics problem for younger students to assess the kinematics of circular motion. Students worked in groups (self-assigned) during the session, and submit their solution at the end of the session.

At the end of the problem-solving session on week 7, we administered a network survey designed on Qualtrics survey service. We distributed the instrument through email to all participants in the three sections.Fig. \ref{survey}, shows the survey design, with question on Column (A) measuring the network of Information Seeking, and question (B) the Network of Good Students. Because the survey is designed upon the roster of participants in the section, we generated a unique survey instrument for each of the three sections.


\subsection{Supplementary Tables}

\clearpage

\subsubsection{Physics Grades}
\begin{table}[!htbp] \centering 
  \caption{OLS multiple regression models for Physics Grades regressed on network predictors.} 
  \label{} 
\scriptsize 
\begin{tabular}{@{\extracolsep{-5pt}}lcccccccc} 
\\[-1.8ex]\hline 
\hline \\[-1.8ex] 
 & \multicolumn{8}{c}{\textit{Dependent variable:}} \\ 
\cline{2-9} 
\\[-1.8ex] & \multicolumn{8}{c}{Physics Grades} \\ 
\\[-1.8ex] & (1) & (2) & (3) & (4) & (5) & (6) & (7) & (8)\\ 
\hline \\[-1.8ex] 
 Mixed &  & 0.849 & 0.895$^{*}$ & 1.179$^{**}$ & 1.229$^{**}$ & 0.543 & 0.665 & 0.993$^{*}$ \\ 
  &  & (0.535) & (0.523) & (0.548) & (0.561) & (0.514) & (0.535) & (0.520) \\ 
  & & & & & & & & \\ 
 Treatment &  & 0.874$^{*}$ & 0.970$^{**}$ & 1.160$^{**}$ & 1.087$^{**}$ & 0.930$^{**}$ & 1.021$^{**}$ & 0.890$^{*}$ \\ 
  &  & (0.484) & (0.475) & (0.494) & (0.485) & (0.458) & (0.490) & (0.467) \\ 
  & & & & & & & & \\ 
 Log(Outdegree) &  &  & $-$0.331$^{*}$ &  &  &  &  &  \\ 
  &  &  & (0.170) &  &  &  &  &  \\ 
  & & & & & & & & \\ 
 Constraint &  &  &  &  & 0.550$^{*}$ &  &  &  \\ 
  &  &  &  &  & (0.315) &  &  &  \\ 
  & & & & & & & & \\ 
 Log(Degree) &  &  &  &  &  & $-$0.394$^{*}$ &  &  \\ 
  &  &  &  &  &  & (0.212) &  &  \\ 
  & & & & & & & & \\ 
 Eigenvector &  &  &  &  &  &  & $-$0.364$^{*}$ &  \\ 
  &  &  &  &  &  &  & (0.206) &  \\ 
  & & & & & & & & \\ 
 P. Elaboration & 0.013 & 0.044 & 0.032 & 0.064 & 0.035 & $-$0.081 & $-$0.047 & $-$0.014 \\ 
  & (0.176) & (0.174) & (0.170) & (0.170) & (0.173) & (0.169) & (0.185) & (0.170) \\ 
  & & & & & & & & \\ 
 Gatekeeper &  &  &  & $-$0.395$^{*}$ &  &  &  &  \\ 
  &  &  &  & (0.201) &  &  &  &  \\ 
  & & & & & & & & \\ 
 Good Student & 0.152 & 0.358 & 0.415 & 0.629$^{**}$ & 0.511$^{*}$ & 0.538$^{**}$ & 0.513$^{*}$ & 0.443$^{*}$ \\ 
  & (0.208) & (0.257) & (0.253) & (0.287) & (0.262) & (0.251) & (0.258) & (0.251) \\ 
  & & & & & & & & \\ 
 UST & 0.422$^{**}$ & 0.572$^{**}$ & 0.656$^{***}$ & 0.601$^{***}$ & 0.655$^{***}$ & 0.608$^{***}$ & 0.554$^{**}$ & 0.600$^{***}$ \\ 
  & (0.208) & (0.224) & (0.223) & (0.220) & (0.240) & (0.212) & (0.217) & (0.217) \\ 
  & & & & & & & & \\ 
 Same City & 0.338 & 0.273 & 0.258 & 0.153 & 0.093 & $-$0.063 & 0.173 & 0.105 \\ 
  & (0.417) & (0.411) & (0.402) & (0.406) & (0.419) & (0.401) & (0.401) & (0.404) \\ 
  & & & & & & & & \\ 
 Eng. Major & $-$0.020 & $-$0.239 & $-$0.434 & $-$0.255 & $-$0.190 & $-$0.097 & $-$0.118 & $-$0.148 \\ 
  & (0.426) & (0.433) & (0.434) & (0.423) & (0.459) & (0.408) & (0.434) & (0.420) \\ 
  & & & & & & & & \\ 
 Female & $-$0.643 & $-$0.515 & $-$0.498 & $-$0.455 & $-$0.484 & $-$0.593 & $-$0.498 & $-$0.586 \\ 
  & (0.400) & (0.400) & (0.391) & (0.392) & (0.397) & (0.376) & (0.387) & (0.387) \\ 
  & & & & & & & & \\ 
 Private School & 0.171 & 0.117 & 0.261 & 0.273 & 0.120 & 0.024 & $-$0.169 & 0.092 \\ 
  & (0.478) & (0.473) & (0.468) & (0.468) & (0.474) & (0.446) & (0.472) & (0.456) \\ 
  & & & & & & & & \\ 
 Mixed*Constraint &  &  &  &  & $-$0.842$^{*}$ &  &  &  \\ 
  &  &  &  &  & (0.422) &  &  &  \\ 
  & & & & & & & & \\ 
 Treatment*Constraint &  &  &  &  & $-$0.052 &  &  &  \\ 
  &  &  &  &  & (0.549) &  &  &  \\ 
  & & & & & & & & \\ 
 Log(Degree)*P.Elab. &  &  &  &  &  & $-$0.363$^{**}$ &  &  \\ 
  &  &  &  &  &  & (0.174) &  &  \\ 
  & & & & & & & & \\ 
 Eigenvector*P.Elab. &  &  &  &  &  &  & $-$0.322$^{*}$ &  \\ 
  &  &  &  &  &  &  & (0.179) &  \\ 
  & & & & & & & & \\ 
 Good Students*P.Elab. &  &  &  &  &  &  &  & $-$0.394$^{**}$ \\ 
  &  &  &  &  &  &  &  & (0.172) \\ 
  & & & & & & & & \\ 
 Constant & 3.875$^{***}$ & 3.626$^{***}$ & 3.606$^{***}$ & 3.416$^{***}$ & 3.696$^{***}$ & 3.907$^{***}$ & 3.746$^{***}$ & 3.702$^{***}$ \\ 
  & (0.509) & (0.519) & (0.507) & (0.518) & (0.517) & (0.496) & (0.523) & (0.502) \\ 
  & & & & & & & & \\ 
\hline \\[-1.8ex] 
Observations & 67 & 67 & 67 & 67 & 67 & 67 & 67 & 67 \\ 
R$^{2}$ & 0.174 & 0.229 & 0.278 & 0.279 & 0.301 & 0.349 & 0.306 & 0.296 \\ 
Adjusted R$^{2}$ & 0.076 & 0.108 & 0.149 & 0.150 & 0.145 & 0.219 & 0.167 & 0.170 \\ 
Residual Std. Error & 1.351 (df = 59) & 1.327 (df = 57) & 1.296 (df = 56) & 1.295 (df = 56) & 1.299 (df = 54) & 1.242 (df = 55) & 1.283 (df = 55) & 1.280 (df = 56) \\ 
F Statistic & 1.779 (df = 7; 59) & 1.886$^{*}$ (df = 9; 57) & 2.160$^{**}$ (df = 10; 56) & 2.168$^{**}$ (df = 10; 56) & 1.934$^{*}$ (df = 12; 54) & 2.685$^{***}$ (df = 11; 55) & 2.200$^{**}$ (df = 11; 55) & 2.349$^{**}$ (df = 10; 56) \\ 
\hline 
\hline \\[-1.8ex] 
\textit{Note:}  & \multicolumn{8}{r}{$^{*}$p$<$0.1; $^{**}$p$<$0.05; $^{***}$p$<$0.01} \\ 
\end{tabular} 
\end{table}

\clearpage
\subsubsection{Problem Elaboration}
\begin{table}[!htbp] \centering 
  \caption{OLS Multiple Regression Models for Problem Elaboration regressed on network predictors.} 
  \label{} 
\scriptsize 
\begin{tabular}{@{\extracolsep{5pt}}lcccccc} 
\\[-1.8ex]\hline 
\hline \\[-1.8ex] 
 & \multicolumn{6}{c}{\textit{Dependent variable:}} \\ 
\cline{2-7} 
\\[-1.8ex] & \multicolumn{6}{c}{Problem Elaboration} \\ 
\\[-1.8ex] & (1) & (2) & (3) & (4) & (5) & (6)\\ 
\hline \\[-1.8ex] 
 Mixed &  & $-$0.355 & $-$0.405 & $-$1.125$^{***}$ & $-$0.222 & $-$0.378 \\ 
  &  & (0.401) & (0.433) & (0.381) & (0.371) & (0.374) \\ 
  & & & & & & \\ 
 Treatment &  & $-$0.127 & $-$0.066 & $-$0.335 & 0.206 & 0.148 \\ 
  &  & (0.365) & (0.377) & (0.324) & (0.350) & (0.353) \\ 
  & & & & & & \\ 
 Constraint &  &  & $-$0.060 &  &  &  \\ 
  &  &  & (0.245) &  &  &  \\ 
  & & & & & & \\ 
 Gatekeeper &  &  &  & 0.588$^{**}$ &  &  \\ 
  &  &  &  & (0.242) &  &  \\ 
  & & & & & & \\ 
 Eigenvector &  &  &  &  & $-$0.456$^{***}$ & $-$0.022 \\ 
  &  &  &  &  & (0.135) & (0.253) \\ 
  & & & & & & \\ 
 Good Student & 0.124 & 0.020 & 0.040 & $-$0.085 & 0.159 & 0.124 \\ 
  & (0.152) & (0.194) & (0.203) & (0.188) & (0.183) & (0.181) \\ 
  & & & & & & \\ 
 UST & $-$0.136 & $-$0.148 & $-$0.217 & $-$0.236 & $-$0.131 & $-$0.186 \\ 
  & (0.151) & (0.168) & (0.184) & (0.149) & (0.155) & (0.155) \\ 
  & & & & & & \\ 
 Same City & $-$0.468 & $-$0.435 & $-$0.515 & $-$0.521$^{*}$ & $-$0.364 & $-$0.324 \\ 
  & (0.300) & (0.305) & (0.318) & (0.271) & (0.282) & (0.279) \\ 
  & & & & & & \\ 
 Eng. Major & $-$0.205 & $-$0.155 & 0.020 & 0.075 & 0.113 & 0.259 \\ 
  & (0.312) & (0.326) & (0.357) & (0.287) & (0.310) & (0.327) \\ 
  & & & & & & \\ 
 Female & 0.307 & 0.287 & 0.240 & $-$0.003 & 0.267 & 0.267 \\ 
  & (0.291) & (0.300) & (0.307) & (0.265) & (0.276) & (0.275) \\ 
  & & & & & & \\ 
 Private School & 0.516 & 0.500 & 0.566 & 0.491 & 0.280 & 0.280 \\ 
  & (0.344) & (0.351) & (0.361) & (0.303) & (0.329) & (0.325) \\ 
  & & & & & & \\ 
 Mixed*Constraint &  &  & 0.190 &  &  &  \\ 
  &  &  & (0.327) &  &  &  \\ 
  & & & & & & \\ 
 Treatment*Constraint &  &  & 0.490 &  &  &  \\ 
  &  &  & (0.422) &  &  &  \\ 
  & & & & & & \\ 
 Mixed*Gatekeeper &  &  &  & $-$2.390$^{***}$ &  &  \\ 
  &  &  &  & (0.466) &  &  \\ 
  & & & & & & \\ 
 Treatment*Gatekeeper &  &  &  & $-$0.521$^{*}$ &  &  \\ 
  &  &  &  & (0.275) &  &  \\ 
  & & & & & & \\ 
 Mixed*Eigenvector &  &  &  &  &  & $-$0.602$^{**}$ \\ 
  &  &  &  &  &  & (0.301) \\ 
  & & & & & & \\ 
 Treatment*Eigenvector &  &  &  &  &  & $-$0.527 \\ 
  &  &  &  &  &  & (0.355) \\ 
  & & & & & & \\ 
 Constant & $-$0.038 & 0.054 & $-$0.042 & 0.151 & $-$0.107 & $-$0.169 \\ 
  & (0.374) & (0.392) & (0.402) & (0.347) & (0.364) & (0.362) \\ 
  & & & & & & \\ 
\hline \\[-1.8ex] 
Observations & 67 & 67 & 67 & 67 & 67 & 67 \\ 
R$^{2}$ & 0.106 & 0.118 & 0.150 & 0.406 & 0.266 & 0.317 \\ 
Adjusted R$^{2}$ & 0.017 & $-$0.003 & $-$0.020 & 0.287 & 0.150 & 0.181 \\ 
Residual Std. Error & 0.991 (df = 60) & 1.002 (df = 58) & 1.010 (df = 55) & 0.844 (df = 55) & 0.922 (df = 57) & 0.905 (df = 55) \\ 
F Statistic & 1.192 (df = 6; 60) & 0.974 (df = 8; 58) & 0.882 (df = 11; 55) & 3.415$^{***}$ (df = 11; 55) & 2.297$^{**}$ (df = 9; 57) & 2.324$^{**}$ (df = 11; 55) \\ 
\hline 
\hline \\[-1.8ex] 
\textit{Note:}  & \multicolumn{6}{r}{$^{*}$p$<$0.1; $^{**}$p$<$0.05; $^{***}$p$<$0.01} \\ 
\end{tabular} 
\end{table}

\end{document}